\documentclass[a4paper]{jpconf}
\usepackage{hyperref}
\usepackage{graphicx}
\usepackage{color}
\usepackage{amsmath,amssymb} % Symbols for itemise
\definecolor{myblue}{rgb}{0.14,0.11,0.49}
\definecolor{myred}{rgb}{0.74,0.22,0.15}
\definecolor{mygreen}{rgb}{0.05,0.52,0.42}
\definecolor{myyellow}{rgb}{0.96,0.92,0.13}
\definecolor{myorange}{rgb}{1,0.61,0.36}
\definecolor{mypurple}{rgb}{0.71,0.02,1}
\definecolor{noir}{gray}{0.} % black

\newcommand{\Couleur}[1]{\textcolor{myblue}{#1}} 
\definecolor{htc}{rgb}{1,1,1} % heading text colour
\newcommand{\Mat}[1]{{{\boldsymbol{#1}}}}
\newcommand{\abs}[1]{\left\vert#1\right\vert}
\def\be{\begin{equation}}
\def\ee{\end{equation}}
\def\bea{\begin{eqnarray}}
\def\eea{\end{eqnarray}}
\def\bc{\begin{center}}
\def\ec{\end{center}}
\def\bi{\begin{itemize}}
\def\ei{\end{itemize}}
\def\bs{\begin{slide}}
\def\es{\end{slide}}
\def\dd{\mathrm{d}}

\def\noi{\noindent}
\begin{document}
%%%%%%%%%%%%%%%%%%%%%%%%%%%%%%%%%%%%%%%%%%%%%%%%%%%%%%%
\title{Progress in evaluating a possible electromagnetic interaction energy in a gravitational field}
\author{Mayeul Arminjon}
\address {Laboratory ``Soils, Solids, Structures, Risks'' (CNRS, Universit\'e Grenoble-Alpes, Grenoble-INP), Grenoble, France}

\ead{Mayeul.Arminjon@3sr-grenoble.fr}

\date{}
			     % (``~`` are extra hard spaces)

\begin{abstract}
\noindent 
The Lorentz-Poincar\'e interpretation of special relativity (SR) keeps the classical concepts of separated space and time, at the price of postulating an indetectable preferred inertial frame or ``ether". But SR does not contain gravity. The presence of gravity could make the ether detectable. This is one idea behind the ``scalar ether theory of gravitation" (SET), which coincides with SR if the gravity field vanishes, and passes a number of tests. However, the coupling of SET with the Maxwell electromagnetic (EM) field needs to use the theory's dynamical equation for the energy tensor in a non-trivial way. It cannot be assumed that the energy tensors of the charged matter and the EM field add to give the total energy tensor, source of the gravitational field. Thus, an additional, ``interaction" energy tensor $\Mat{T}_\mathrm{inter}$ has to be postulated. Asking that $\Mat{T}_\mathrm{inter}$ is Lorentz-invariant in the situation of SR, fixes its form. It depends only on a scalar field $p$. $\Mat{T}_\mathrm{inter}$ is an exotic kind of matter and is distributed in the whole space, hence it could contribute to dark matter. For a weak gravitational field, $p$ obeys a first-order partial differential equation (PDE) involving the EM field and the Newtonian potential. However, the EM field varies on the scale of the wavelength, which is extremely small. To get the field $p$ in a galaxy, some averaging has to be done. After several attempts based on the homogenization theory, a simpler way has been found recently: If the macro-averages of $p$ and the EM field vary smoothly, it can be shown that the PDE for $p$ remains valid in the same form with spacetime-averaged fields. The current stage of calculations will also been shown.
\end{abstract}
%%%%%%%%%%%%%%%%%%%%%%%%%%%%%%%%%%%%%%%%%%%%%%%%%%%%%%%%%%%%%%%%%%%%%%%%%%%
%%%%%%%%%%%%%%%%%%%%%%%%%%%%%%%%%%%%%%%%%%%%%%%%%%%%%%%%%%%%%%%%%%%%%%%%%%%%%%%%
\section{Introduction}

In this paper, we advance the test of a special consequence of an alternative theory of gravitation. Namely, this theory says that, in the presence of both a gravity field and an electromagnetic field, an additional exotic energy tensor appears, that could contribute to dark matter. We begin with presenting one important motivation of that theory and briefly summarizing the theory. In special relativity (SR), the time intervals between two events, and even the notion of simultaneous events, become dependent on the reference frame that is considered. Therefore, in its standard interpretation, introduced in the celebrated papers of Einstein \cite{Einstein1905} and Minkowski \cite{Minkowski1907}, SR is thought to enforce us to abandon the classical concepts of separated space and time. However, there is another interpretation of SR, initiated by Lorentz \cite{Lorentz1904} and Poincar\'e \cite{Poincare1905, Poincare1906}, which sees the space contraction and time dilation as absolute effects of motion through a preferred inertial frame or ``ether". This alternative interpretation, which may be called the Lorentz-Poincar\'e interpretation of SR, has been discussed by a number of authors. In particular, Prokhovnik \cite{Prokhovnik1967} has made a very detailed comparison between the Einstein-Minkowski and Lorentz-Poincar\'e interpretations of SR, and has proved that these two interpretations are physically exactly equivalent, while corresponding to two different logics: the ``logic of spacetime" vs. the ``logic of absolute motion". 
\footnote{\
Actually, there is a physical difference between the two interpretations, which has not been envisaged by Prokhovnik nor, it seems, by other researchers discussing the Lorentz-Poincar\'e interpretation: since, in the latter interpretation, the only ``true time" is the synchronized time in the ether reference frame, superluminal signals would not lead to a nonsense in this version, and hence are conceivable \cite{O3}.
}
In the Lorentz-Poincar\'e interpretation of SR, the inertial time of the preferred inertial frame (ether), synchronized using the Poincar\'e-Einstein synchronization convention, is thought of as the ``true time", and the simultaneity thus defined in the ether frame is thought of as the ``absolute simultaneity". The reciprocity of the metrical effects of length contraction and time dilation (when interchanging the respective roles of two inertial frames) is then seen as an illusion due to the operational necessity of still using the Poincar\'e-Einstein synchronization convention in a moving frame. Yet because the Lorentz-Poincar\'e theory is really equivalent to SR, that ether is indetectable and all inertial frames are equally good candidates for being the ether!\\

But SR does not include gravity. Gravity might violate relativity and reveal the ether. (Note that even general relativity (GR) does not obey SR's relativity principle, since there are no global inertial frames and no global Lorentz transformations in GR \cite{Fock1964}.) Such a violation does happen in the ``scalar ether theory" of gravitation (SET), which starts from gravity as being Archimedes' thrust in a fluid ether filling the space \cite{O3}. SET coincides with SR when the gravitational field vanishes. It endows spacetime with two metrics: a flat ``background" metric \Couleur{$\Mat{\gamma}^0$} and a curved ``physical" metric \Couleur{$\Mat{\gamma}$}. Motion is defined by an extension of Newton's second law to a curved spacetime, which implies that free particles follow the geodesics of \Couleur{$\Mat{\gamma}$} only for a static gravitational field (see Ref. \cite{O3} and references therein). SET passes a number of tests. See Ref. \cite{A54}, Sect. 1, for a summary: in particular, all gravitational effects on light rays are identical, to the relevant post-Newtonian approximation, with the standard predictions of GR. See Ref. \cite{A59} for results regarding celestial mechanics: ephemerides nearly coincident with those of GR are obtained, although the question of the absolute velocity of the solar system in that theory remains open.\\

However, in that theory, the modification of electrodynamics due to the presence of a gravitational field is not trivial. We will first briefly recall (Sect. \ref{ED_in_SET}) why this is the case and why, as announced at the beginning of this paper, that modification leads us to introduce an additional energy tensor. Then in Sect. \ref{InteractEqn} we will show the equations that should determine that energy tensor. In Sect. \ref{Averaging} we will describe the path that we had to follow in order to find an appropriate averaging of the fields and the equation. Section \ref{Solving-the-PDE} shows the current state of numerically implementing the solution of the averaged equation. Finally, Sect. \ref{Conclusion} presents our conclusions. The new part in this work begins with \S \ref{ST-homogenizn}.
%%%%%%%%%%%%%%%%%%%%%%%%%%%%%%%%%%%%%%%%%%%%%%%%%%%%%%%%%%%%%%%%%%%%%%%%%%%%%%%%

%%%%%%%%%%%%%%%%%%%%%%%%%%%%%%%%%%%%%%%%%%%%%%%%%%%%%%%%%%%%%%%%%%%%%%%%%%%%%%%%

%%%%%%%%%%%%%%%%%%%%%%%%%%%%%%%%%%%%%%%%%%%%%%%%%%%%%%%%%%%%%%%%%%%%%%%%%%%%%%%%
\section{Electrodynamics in SET and the interaction energy tensor}\label{ED_in_SET}
%%%%%%%%%%%%%%%%%%%%%%%%%%%%%%%%%%%%%%%%%%%%%%%%%%%%%%%%%%%%%%%%%%%%%%%%%%%%%%%%

%%%%%%%%%%%%%%%%%%%%%%%%%%%%%%%%%%%%%%%%%%%%%%%%%%%%%%%%%%%%%%%%%%%%%%%%%%%%%%%%
\subsection{Electrodynamics in the presence of gravity in SET}
%%%%%%%%%%%%%%%%%%%%%%%%%%%%%%%%%%%%%%%%%%%%%%%%%%%%%%%%%%%%%%%%%%%%%%%%%%%%%%%%

The equations of electrodynamics of GR rewrite those of SR by using the ``comma goes to semicolon" rule:\Couleur{$\quad _{,\,\nu} \ \rightarrow \ _{;\,\nu}$}, i.e., partial derivatives are replaced by covariant derivatives. This is not possible in SET, because the dynamical equation for the energy(-momentum-stress) tensor \Couleur{$\Mat{T}$} is not generally \Couleur{$ T^{\lambda \nu }_{\ \, ;\nu}=0$} (which rewrites \Couleur{$T^{\lambda \nu }_{\ \, ,\nu}=0$} valid in SR). In SET, the first set of Maxwell equations remains valid \cite{A54}. The second set (the set of the four so-called ``inhomogenous" Maxwell equations) is deduced from the two sets of dynamical equations verified by (i) the total energy tensor and (ii) the energy tensor of the charged medium in the presence of the Lorentz force. In a first version, it was assumed that (as is the case in GR):
\be
\hypertarget{Ass1}{\mathrm{(A)\ \ Total\ energy\ tensor\ }} \Couleur{\Mat{T} = \Mat{T}_\mathrm{charged\ medium} + \Mat{T}_\mathrm{field}}. 
\ee
The additivity \hyperlink{Ass1}{(A)} leads to a form of Maxwell's second group in SET \cite{A54}. But that form predicts charge production/destruction at untenable rates, and hence had to be discarded \cite{A56}.\\

The additivity assumption \hyperlink{Ass1}{(A)} is contingent and may be abandoned. It means introducing an ``interaction" energy tensor \Couleur{$\Mat{T}_\mathrm{inter}$} such that, in contrast with \hyperlink{Ass1}{(A)} that applies in GR, we have
\be\label{Tinter}
\Couleur{\Mat{T}_{(\mathrm{total})} = \Mat{T}_\mathrm{charged\ medium} + \Mat{T}_\mathrm{field}\ \underline{+ \Mat{T}_\mathrm{inter}}\,}. 
\ee
One then has to constrain the form of \Couleur{$\Mat{T}_\mathrm{inter}$} and derive equations for it.

%%%%%%%%%%%%%%%%%%%%%%%%%%%%%%%%%%%%%%%%%%%%%%%%%%%%%%%%%%%%%%%%%%%%%%%%%%%%%%%%
\subsection{Form of the interaction energy tensor}
%%%%%%%%%%%%%%%%%%%%%%%%%%%%%%%%%%%%%%%%%%%%%%%%%%%%%%%%%%%%%%%%%%%%%%%%%%%%%%%%

In SR, the additivity assumption \hyperlink{Ass1}{(A)} holds, thus \Couleur{$\Mat{T}_\mathrm{inter}={\bf 0}$}. In SET we may impose that \Couleur{$\Mat{T}_\mathrm{inter}$} should be Lorentz-invariant in the situation of SR, i.e. when the metric \Couleur{$\Mat{\gamma}$} is Minkowski's metric \Couleur{$\Mat{\gamma}^0$} \ (\Couleur{$\gamma^0 _{\mu \nu }=\eta _{\mu \nu }$} in Cartesian coordinates). This leads uniquely \cite{A57} to the following definition:
\be\label{T_inter}
\Couleur{(T_\mathrm{inter})^\mu\ {_\nu }:=  p\,\delta ^\mu _\nu}, \qquad \mathrm{or}\quad \Couleur{(T_\mathrm{inter})^{\mu  \nu }:=  p\,\gamma^{\mu \nu}},
\ee
with some scalar field \Couleur{$p$}. The corresponding interaction energy is
\be\label{E_inter}
\Couleur{E_\mathrm{inter}:=T^{0 0}_\mathrm{inter} = p\gamma ^{0 0}}.
\ee
The medium with energy tensor (\ref{T_inter}) can be counted as ``dark matter", because:
\bi
\item It is not localized inside usual matter: we have \Couleur{$ p\ne 0$} at a generic point.
\item It is gravitationally active, for \Couleur{$T^{0 0}\,$} is the source of the gravitational field in SET \cite{A59}.
\item It is not usual matter: e.g., no velocity can be defined in view of the Lorentz invariance of the tensor (\ref{T_inter}). (The invariance is ``pointwise".)
\ei 

%%%%%%%%%%%%%%%%%%%%%%%%%%%%%%%%%%%%%%%%%%%%%%%%%%%%%%%%%%%%%%%%%%%%%%%%%%%%%%%%
\subsection{Dynamical equations and the second Maxwell set}\label{2nd-Maxwell}
%%%%%%%%%%%%%%%%%%%%%%%%%%%%%%%%%%%%%%%%%%%%%%%%%%%%%%%%%%%%%%%%%%%%%%%%%%%%%%%%
With the general decomposition (\ref{Tinter}), the second set of Maxwell equations is still deduced from the two sets of dynamical equations verified by (i) the total energy tensor:
\be\label{DT-SET}
\Couleur{T^{\mu  \nu}_{\ \ ;\nu} =b^\mu (\Mat{T})},
\ee
and (ii) the energy tensor of the charged medium:
\be\label{Dyn_T-chg-SET}
\Couleur{T^{\mu  \nu}_{\mathrm{charged\ medium}\ \ ;\nu} =b^\mu (\Mat{T}_\mathrm{charged\ medium})+ F^\mu_{\mathrm{exact}\  \nu  }\,J^\nu}.
\ee
Here \Couleur{$F^\mu_{\mathrm{exact}\  \nu  }$} are the mixed components of the Faraday field tensor \Couleur{$\Mat{F}_\mathrm{exact}$}, \Couleur{$J^\mu$} is the 4-current, and
\be\label{b^mu}
\Couleur{b^0(\Mat{T}) := \frac{1}{2}\,\gamma^{00}\,g_{ij,0}\,T^{ij}},
\quad \Couleur{b^i(\Mat{T}) := \frac{1}{2}\,g^{ij}\,g_{jk,0}\,T^{0k}},
\ee
with \Couleur{$g_{ij}$} the components of the spatial metric tensor \Couleur{$\Mat{g}=\Mat{g}_\mathcal{E}$} associated \cite{L&L, Moller1952, A54} with the spacetime metric \Couleur{$\Mat{\gamma }$} in the preferred reference frame \Couleur{$\mathcal{E}$}, and \Couleur{$(g^{ij})$} the inverse matrix of \Couleur{$(g_{ij})$}. (These equations are covariant only under coordinates changes that are internal to the frame \Couleur{$\mathcal{E}$}.) By combining Eqs. (\ref{T_inter}), (\ref{DT-SET}), and (\ref{Dyn_T-chg-SET}), we indeed obtain \cite{A57}:
\be\label{Maxwell 2 SET p}
\Couleur{F^{\mu  \nu }_{\mathrm{exact}\ \ ;\nu }} = \Couleur{\mu _0 \left[ G^\mu_{\mathrm{exact}\ \nu  } \left (b^\nu (\Mat{T}_\mathrm{field})-\delta ^\nu (p)\right ) -J^\mu \right ]},
\ee
where %\Couleur{$F^{\mu  \nu }_\mathrm{exact}$} are the contravariant components of \Couleur{$\Mat{F}_\mathrm{exact}$}, 
\Couleur{$(G^\mu_{\mathrm{exact}\  \nu  })$} is the inverse of the matrix \Couleur{$(F^\mu_{\mathrm{exact}\  \nu  })$}, and \Couleur{$\delta ^\nu (p)$} depends, in addition to the scalar field \Couleur{$p$} defining the interaction tensor (\ref{T_inter}), on the other scalar field \Couleur{$\beta = \sqrt{\gamma_{0 0}}$} (in coordinates adapted to the preferred reference frame, including a preferred time coordinate \Couleur{$T$}), which is the scalar gravitational field of SET.

%%%%%%%%%%%%%%%%%%%%%%%%%%%%%%%%%%%%%%%%%%%%%%%%%%%%%%%%%%%%%%%%%%%%%%%%%%%%%%%%
\section{Charge conservation and the equation for the interaction energy tensor}\label{InteractEqn}
%%%%%%%%%%%%%%%%%%%%%%%%%%%%%%%%%%%%%%%%%%%%%%%%%%%%%%%%%%%%%%%%%%%%%%%%%%%%%%%%

By using the identity \Couleur{$F^{\mu \nu } _{\ \ \,;\nu;\mu   }\equiv 0$}, we get from (\ref{Maxwell 2 SET p}):
\be\label{divJ SET p}
\Couleur{J^\mu_{\ \, ;\mu} = \left[ G^\mu_{\mathrm{exact}\ \nu  } \left (b^\nu (\Mat{T}_\mathrm{field})-\delta ^\nu (p)\right ) \right ]_{;\mu}}.
\ee
With the interaction energy tensor (\ref{T_inter}) we have just one unknown more as compared with the case with \hyperlink{Ass1}{additive energy tensors}: the scalar field \Couleur{$p$}. So we need just one scalar equation more. As is clear from (\ref{divJ SET p}), we may add {\it charge conservation} as the new scalar equation, thus equating the r.h.s. of (\ref{divJ SET p}) to zero:
\be\label{Jmumu=0}
\Couleur{\left[ G^\mu_{\mathrm{exact}\ \nu  } \left (b^\nu (\Mat{T}_\mathrm{field})-\delta ^\nu (p)\right ) \right ]_{;\mu} = 0}.
\ee
Then the system of equations of electrodynamics of SET is again closed, and satisfies charge conservation. To simplify Eq. (\ref{Jmumu=0}), we consider the relevant case of a given weak and slowly varying gravitational field with Newtonian potential \Couleur{$U$} \cite{A57}. That assumption allows us to introduce asymptotic expansions of the fields: the gravitational field, the EM field (although the latter is not assumed weak nor slowly varying) \cite{A56}, and also the field $p$ \cite{A57}. Then the charge conservation equation (\ref{Jmumu=0}) rewrites as the following PDE for the first approximation of \Couleur{$p$} (which here we shall still denote by \Couleur{$p$}) \cite{A57}:
\be\label{Eq for p}
\fbox{$\Couleur{\mathrm{div}_4 \left (\Mat{G}.\nabla_4p\right):= \left (G^{\mu \nu }\, p_{,\nu}\right )_{,\mu}= f}.$}
\ee
Here, the \Couleur{$G^{\mu \nu }$}'s are the components of an antisymmetric spacetime tensor \Couleur{$\Mat{G}$}: the inverse tensor of the Faraday field tensor of the first approximation, that obeys the flat-spacetime Maxwell equations. In addition, in Eq. (\ref{Eq for p}), we have
\be
\fbox{$\Couleur{f: = \left( d^i \partial _T U \right)_{,i }},$}
\ee
with \Couleur{$d^i\ (i=1,2,3)$} the components of a spatial vector \Couleur{${\bf d}$} made with \Couleur{$({\bf E},{\bf B})$}. The time derivative \Couleur{$\partial _T U $} has to be taken in the preferred reference frame and with the preferred time coordinate \Couleur{$T$}.

%%%%%%%%%%%%%%%%%%%%%%%%%%%%%%%%%%%%%%%%%%%%%%%%%%%%%%%%%%%%%%%%%%%%%%%%%%%%%%%%
\section{Averaging the PDE for $\Mat{T}_\mathrm{inter}$}\label{Averaging}
%%%%%%%%%%%%%%%%%%%%%%%%%%%%%%%%%%%%%%%%%%%%%%%%%%%%%%%%%%%%%%%%%%%%%%%%%%%%%%%%

In order to check if the interaction energy (\ref{E_inter}) may contribute significantly to dark matter, we need to integrate {\it on a galactic scale} \Couleur{$r \sim 10^{19} \mathrm{m}$} the PDE (\ref{Eq for p}) for the scalar field \Couleur{$p$}. However, the given fields \Couleur{$\Mat{G}$} and \Couleur{$f$} in (\ref{Eq for p}) oscillate on the scale \Couleur{$r\sim \lambda \simeq 10^{-6}\mathrm{m}$} and \Couleur{$t\sim \lambda/c$}, as do \Couleur{${\bf E}$} and \Couleur{${\bf B}$}. Therefore, we have no chance to succeed in the integration! This situation is typical of the {\it homogenization theory}. The aim of that theory is to get ``homogenized" PDEs allowing one to describe at the macroscopic scale the medium, assumed periodic or quasi-periodic at a microscopic scale. For Eq. (\ref{Eq for p}), the ``medium" is characterized by the pair of given heterogeneous fields \Couleur{$(\Mat{G},f)$}. In \S\S 4.1 and 4.2, we will summarize our past attempt \cite{B43} to use that theory to obtain a tractable PDE for some relevant average of the unknown scalar field \Couleur{$p$}; in \S \ref{ST-homogenizn} we will present the later developments of this approach, in the ``spacetime homogenization" setting. Then in \S\S \ref{ST-averaging} to \ref{AveragedPDE} we will present and develop a simpler approach, the ``spacetime averaging", that turns out to be appropriate for the present problem.

%%%%%%%%%%%%%%%%%%%%%%%%%%%%%%%%%%%%%%%%%%%%%%%%%%%%%%%%%%%%%%%%%%%%%%%%%%%%%%%%
\subsection{A bit more on homogenization theory}
%%%%%%%%%%%%%%%%%%%%%%%%%%%%%%%%%%%%%%%%%%%%%%%%%%%%%%%%%%%%%%%%%%%%%%%%%%%%%%%%

The homogenization theory considers two spacetime variables related by a small parameter \Couleur{$\epsilon \ll 1$}: 
\bi
\item A slow variable \Couleur{${\bf X} $}, that browses the medium at macroscopic scale.

\item A quick variable, \Couleur{${\bf Y} = {\bf X}/\epsilon$}: an $O(1)$ variation of it browses the quasi-period of the medium.\\
\ei
The fields are stated to be functions of \Couleur{${\bf X}$} {\it and} \Couleur{${\bf Y}$}, {\it periodic} or quasi-periodic with respect to \Couleur{${\bf Y}$}. {\it Asymptotic expansions} are stated, e.g.
\be\label{p expansion-spacetime}
\Couleur{p^\epsilon({\bf X}) = p_0\left({\bf X},{\bf Y}\right) \epsilon^0 + p_1\left({\bf X},{\bf Y}\right) \epsilon + O(\epsilon^2)} , \qquad \Couleur{{\bf Y} =\frac{{\bf X}}{\epsilon}}
\ee

%%%%%%%%%%%%%%%%%%%%%%%%%%%%%%%%%%%%%%%%%%%%%%%%%%%%%%%%%%%%%%%%%%%%%%%%%%%%%%%%
\subsection{Homogenizing the PDE for $\Mat{T}_\mathrm{inter}$: the different possible ways}
%%%%%%%%%%%%%%%%%%%%%%%%%%%%%%%%%%%%%%%%%%%%%%%%%%%%%%%%%%%%%%%%%%%%%%%%%%%%%%%%

Depending on which spacetime variable is considered primordial, there are
three possibilities:\\

\bi
\item Time Homogenization: The homogenization theory applies quite straightforwardly \cite{B43}. But the remaining space dependence at scale \Couleur{$r\sim \lambda \simeq 10^{-6}\mathrm{m}$} prevents integration of the PDE at galactic scale.\\

\item Space Homogenization: The homogenization theory applies less well \cite{B43}. Anyway, the remaining time dependence at scale \Couleur{$t\sim \lambda/c$} also prevents integration of the PDE at a galactic scale.\\

\item Spacetime Homogenization.

\ei

%%%%%%%%%%%%%%%%%%%%%%%%%%%%%%%%%%%%%%%%%%%%%%%%%%%%%%%%%%%%%%%%%%%%%%%%%%%%%%%%
\subsection{Spacetime homogenization} \label{ST-homogenizn}
%%%%%%%%%%%%%%%%%%%%%%%%%%%%%%%%%%%%%%%%%%%%%%%%%%%%%%%%%%%%%%%%%%%%%%%%%%%%%%%%

The PDE (\ref{Eq for p}) for \Couleur{$p$} has just the same form as the {\it stationary heat conduction equation for the temperature}  \Couleur{$\theta $}, except that here we have 4-d {\it spacetime} instead of 3-d space. Therefore, we may adapt known results \cite{Caillerie2012}, despite the important difference that here the spacetime tensor \Couleur{$\Mat{G}$} is antisymmetric, whereas the conductivity (spatial) tensor \Couleur{$\Mat{K}$} is symmetric. The main result \cite{B43} is that the homogenized PDE has the same form as (\ref{Eq for p}), replacing \Couleur{$\Mat{G}$} by a ``homogenized" tensor \Couleur{$\Mat{G}^\mathrm{H}$}. However, \Couleur{$\Mat{G}^\mathrm{H}$} is {\it not} the local spacetime average of the ``microscopic" tensor \Couleur{$\Mat{G}$}: \Couleur{$\Mat{G}^\mathrm{H}$} is obtained by solving a boundary value problem on a local microscopic cell for a linear first-order PDE. Adapting Ref. \cite{Caillerie2012}, we find that here this problem is 
\be
\Couleur{k^\mu \,\chi^\nu_{\ ,\,\mu} = -k^\nu  \quad (\nu =0,...,3), \qquad k^\nu := G^{\mu \nu }_{\ ,\,\mu}}
\ee
with periodic boundary conditions on the elementary cell in spacetime. 
\footnote{\
Note that \Couleur{$ k^\nu $} is a 4-vector, though in general coordinates the definition above should be rewritten as \Couleur{$k^\nu := G^{\mu \nu }_{\ | \,\mu}$} with \Couleur{$| $} the covariant derivatives relative to the flat ``background" metric \Couleur{$\Mat{\gamma}^0$}. However, we calculate \Couleur{$ k^\nu $} in an inertial frame for \Couleur{$\Mat{\gamma}^0$}, hence those covariant derivatives coincide with the partial derivatives.
}
This problem has to be solved by the finite element method. This has been successfully numerically implemented, by using the freeware \verb+FreeFem+. However, after having solved this problem, it remains to solve \Couleur{$\left (G^{\mathrm{H} \ {\mu \nu }}\, P_{,\nu}\right )_{,\mu}= F$} for the unknown \Couleur{$P : = \langle p \rangle$} and with data \Couleur{$F : = \langle f \rangle$}. (Here \Couleur{$ \langle f \rangle$} means the volume average in the elementary cell in spacetime, that depends on the macroscopic point.) Thus this method is very heavy. Fortunately we can now present another method, that is much simpler.

%%%%%%%%%%%%%%%%%%%%%%%%%%%%%%%%%%%%%%%%%%%%%%%%%%%%%%%%%%%%%%%%%%%%%%%%%%%%%%%%
\subsection{Spacetime averaging}\label{ST-averaging}
%%%%%%%%%%%%%%%%%%%%%%%%%%%%%%%%%%%%%%%%%%%%%%%%%%%%%%%%%%%%%%%%%%%%%%%%%%%%%%%%

Because \Couleur{$G^{\mu \nu } = -G^{\nu \mu }$}, Eq. (\ref{Eq for p}) rewrites as a {\it first-order} PDE:
\be\label{1st_order_Eq_for_p}
\fbox{$\Couleur{k^\nu p_{,\nu} = f},$} \qquad \mathrm{or}\quad \Couleur{{\bf k.}\nabla p = f \qquad  (\nabla:=\nabla_4)}.
\ee
And again because \Couleur{$G^{\mu \nu } = -G^{\nu \mu }$}, we have
\be\label{div k = 0}
\Couleur{\mathrm{div}\, {\bf k}:=\mathrm{div}_4 \, {\bf k}:= k^\nu_{,\,\nu} := G^{\mu \nu }_{\ ,\,\mu\,,\nu}=0.}
\ee
Recall that all fields here vary with pseudo-periods \Couleur{$\lambda \simeq 1 \mu \mathrm{m}$} and \Couleur{$T \simeq \lambda/c$}, which are {\it extremely} small with respect to galactic scales of space and time. We assume that the fields \Couleur{$ {\bf k}$} and \Couleur{$ \nabla p$} are ``locally macro-homogeneous". I.e., they are {\it slow variations} of {\it macro-homogeneous fields}. In brief, the latter means that the volume averages \Couleur{$ {\bf K} = \overline{{\bf k}}$} and \Couleur{$ \overline{\nabla p } $} are stationary. Details follow.

%%%%%%%%%%%%%%%%%%%%%%%%%%%%%%%%%%%%%%%%%%%%%%%%%%%%%%%%%%%%%%%%%%%%%%%%%%%%%%%%
\subsection{Macro-homogeneous fields \Couleur{$ {\bf k}$} and \Couleur{$\nabla p$} }
%%%%%%%%%%%%%%%%%%%%%%%%%%%%%%%%%%%%%%%%%%%%%%%%%%%%%%%%%%%%%%%%%%%%%%%%%%%%%%%%

We provisionally forget the slow variation of the fields. We formulate precise definitions of macro-homogeneous fields \Couleur{${\bf k}$} and \Couleur{$\nabla p$}, i.e., respectively \hyperlink{Ass(i)}{(i)} and \hyperlink{Ass(ii)}{(ii)} below; and we state a sufficient validity condition for a ``no-correlation condition" between them. This has some similarity with Ref. \cite{A6}, Sect. 2, in which the role of the 4-vector \Couleur{${\bf k}$} here is played by the stress (a spatial second-order tensor), while the role of the scalar \Couleur{$p$} here is played by the velocity (a spatial vector). We state the following\\
 
\noi \hypertarget{Proposition}{{\bf Proposition.}} {\it Assume that:}
\hypertarget{Ass(i)}{(i)} \Couleur{$ {\bf k} = {\bf k}_0 + \delta {\bf k}$} {\it with} \Couleur{${\bf k}_0$} {\it constant,} \Couleur{$ {\bf \delta k}$} {\it bounded, and, for (hyper)cubes} \Couleur{$\Omega$} in \Couleur{$\mathbb{R}^4$}{\it , with side} \Couleur{$R(\Omega)$} {\it and (hyper)volume} \Couleur{$V(\Omega)= R(\Omega )^4$} ,

\be\label{int delta _k ->0}
\Couleur{\frac{1}{V(\Omega)} \int_\Omega {\bf \delta k} \, \dd V \rightarrow {\bf 0}} \quad \mathrm{as}\quad \Couleur{R(\Omega) \rightarrow \infty},
\ee
{\it independently of the position of the cubes.}
\vspace{2mm}
\hypertarget{Ass(ii)}{(ii)}\ \Couleur{$p({\bf X}) = {\bf g}_0 {\bf .X} + \delta p$}, {\it with (}\Couleur{$\partial \Omega$} {\it being the boundary of the cube} \Couleur{$\Omega$} {\it and} \Couleur{$\dd S$} {\it being the (hyper)surface element on} \Couleur{$\partial \Omega$}{\it ):}

\be\label{int |delta p| dS -> 0}
\Couleur{\frac{1}{V(\Omega)} \int_{\partial \Omega} \vert \delta p \vert \, \dd S \rightarrow 0} \quad \mathrm{as}\quad \Couleur{R(\Omega) \rightarrow \infty}.
\ee
\vspace{2mm}
\noi \hypertarget{Ass(iii)}{(iii)} \Couleur{$\mathrm{div}\, {\bf k} = 0$}. [Note that this is always true for the relevant field \Couleur{$ {\bf k} $}, Eq. (\ref{div k = 0}).]\\

\noi {\it Define for a general field \Couleur{$h$} the volume average in \Couleur{$\Omega $}:}
\be\label{bar Omega}
\Couleur{\overline{h}^{\ \Omega} := \frac{1}{V(\Omega)} \int_\Omega h\, \dd V },
\ee
{\it and set}
\be\label{Delta_Omega}
\Couleur{\Delta _\Omega := \overline{{\bf k.}\nabla p}^{\ \Omega} - \overline{{\bf k}}^{\ \Omega} \,{\bf .}\,\overline{\nabla p}^{\ \Omega} }.
\ee

\noi {\it Then, we have}
\be\label{kbar -> k0}
\Couleur{\overline{{\bf k}}^{\ \Omega} \rightarrow {\bf k}_0} \quad \mathrm{as}\quad \Couleur{R(\Omega) \rightarrow \infty},
\ee
\be\label{grad-p bar -> g0}
\Couleur{ \overline{\nabla p }^{\ \Omega}  \rightarrow {\bf g}_0} \quad \mathrm{as}\quad \Couleur{R(\Omega) \rightarrow \infty},
\ee
\be\label{Delta_Omega -> 0}
\Couleur{\Delta _\Omega \rightarrow 0 } \quad \mathrm{as}\quad \Couleur{R(\Omega) \rightarrow \infty}.
\ee

\vspace{2mm}
\noindent {\it Proof.} It is immediate to see that \Couleur{$ {\bf k} = {\bf k}_0 + \delta {\bf k}$} together with (\ref{int delta _k ->0}) %assumption \hyperlink{Ass(i)}{(i)} 
implies (\ref{kbar -> k0}). On the other hand, by using the divergence theorem, one checks easily the (well-known) formula
\be\label{int grad p dV = int pn dS}
\Couleur{\int_\Omega \nabla p \, \dd V = \int_{\partial \Omega} p \,{\bf n}\, \dd S},
\ee 
where \Couleur{${\bf n}$} is the exterior normal. By applying this to \Couleur{$\delta p$} in the place of \Couleur{$p$}:
\be\label{int grad delta p dV = int delta pn dS}
\Couleur{\int_\Omega \nabla \delta p \, \dd V = \int_{\partial \Omega} \delta p \, {\bf n}\, \dd S},
\ee 
it follows that assumption \hyperlink{Ass(ii)}{(ii)}  implies (\ref{grad-p bar -> g0}). Also, if assumption \hyperlink{Ass(iii)}{(iii)} is valid (\Couleur{$\mathrm{div}\, {\bf k} = 0$}), we have 
\be\label{k.grad p}
\Couleur{{\bf k.}\nabla p = \mathrm{div}\, ({\bf k} p)},
\ee
hence using again the divergence theorem:
\be\label{k.grad p bar}
\Couleur{\overline{{\bf k.}\nabla p}^{\ \Omega}} = \Couleur{\frac{1}{V(\Omega)} \int_{\partial \Omega} {\bf k.n}\,p\, \dd \, S}.
\ee
Defining
\be\label{Delta'_Omega}
\Couleur{\Delta '_\Omega := \overline{({\bf k}-{\bf k}_0){\bf .}(\nabla p - {\bf g}_0)}^{\ \Omega} = \overline{\delta {\bf k}{\bf .}\nabla \delta p }^{\ \Omega}} ,
\ee
we note that \Couleur{$\mathrm{div}\, {\bf k} = \mathrm{div}\, \delta{\bf k}$} by assumption \hyperlink{Ass(i)}{(i)}, hence \Couleur{$\mathrm{div}\, \delta{\bf k} =0$} if assumption \hyperlink{Ass(iii)}{(iii)} is valid, so that we may apply (\ref{k.grad p bar}) to \Couleur{$\delta{\bf k}$} and \Couleur{$\delta p$} in the place of \Couleur{${\bf k}$} and \Couleur{$ p$}:
\be\label{dk.grad dp bar}
\Couleur{\Delta '_\Omega = \Couleur{\frac{1}{V(\Omega)} \int_{\partial \Omega} \delta {\bf k}\,{\bf.n}\,\delta p\, \dd \, S}}.
\ee
Therefore, if in accordance with assumption \hyperlink{Ass(i)}{(i)} \Couleur{$\delta {\bf k}$} is bounded: \Couleur{$\abs{\delta {\bf k}} \le M$}, and if assumption \hyperlink{Ass(ii)}{(ii)} is valid, we obtain:
\be\label{Delta'_Omega->0}
\Couleur{\abs{\Delta '_\Omega} \le \frac{M}{V(\Omega)} \int_{\partial \Omega} \abs{\delta p}\, \dd \, S \rightarrow 0} \quad \mathrm{as}\quad \Couleur{R(\Omega) \rightarrow \infty}.
\ee
Finally, we have from the definitions (\ref{Delta_Omega}) and (\ref{Delta'_Omega}):
\be
\Couleur{\Delta'_\Omega - \Delta_\Omega = \left (\overline{{\bf k}}^{\ \Omega} - {\bf k}_0 \right ) {\bf .} \overline{\nabla p}^{\ \Omega} + \left ({\bf k}_0-\overline{{\bf k}}^{\ \Omega}\right ) {\bf .g}_0}.
\ee
Hence, if assumptions \hyperlink{Ass(i)}{(i)} to \hyperlink{Ass(iii)}{(iii)} hold, we obtain by using (\ref{kbar -> k0}) and (\ref{grad-p bar -> g0}):
\be
\Couleur{\Delta'_\Omega - \Delta_\Omega \rightarrow 0} \quad \mathrm{as}\quad \Couleur{R(\Omega) \rightarrow \infty}.
\ee
Together with (\ref{Delta'_Omega->0}), this proves (\ref{Delta_Omega -> 0}). This completes the proof of the \hyperlink{Proposition}{Proposition}. \hfill $\square$\\

{\it Remark.} One may observe some dissymmetry in the assumptions for the macro-homogeneity of \Couleur{$ {\bf k}$} (\hyperlink{Ass(i)}{Assumption (i)}) and \Couleur{$\nabla p$} (\hyperlink{Ass(ii)}{Assumption (ii)}). This dissymmetry is due to the wish to ensure the no-correlation condition (\ref{Delta_Omega -> 0}). Indeed, to obtain (\ref{kbar -> k0}), one uses (\ref{int delta _k ->0}), but one does not use the assumption that \Couleur{$ {\bf \delta k}$} is bounded. Also, to obtain (\ref{grad-p bar -> g0}), one may replace (\ref{int |delta p| dS -> 0}) by the weaker assumption 
\be\label{int delta p dS -> 0}
\Couleur{\frac{1}{V(\Omega)} \int_{\partial \Omega} \delta p \, {\bf n} \, \dd S \rightarrow 0} \quad \mathrm{as}\quad \Couleur{R(\Omega) \rightarrow \infty},
\ee
as one sees from (\ref{int grad delta p dV = int delta pn dS}). And, again due to (\ref{int grad delta p dV = int delta pn dS}), assuming (\ref{int delta p dS -> 0}) is the same for \Couleur{$\nabla p$} as is assuming (\ref{int delta _k ->0}) for \Couleur{${\bf k}$}. %Thus, to obtain the stationarity of the volume averages, .....
However, to prove that \Couleur{$\Delta'_\Omega \rightarrow 0 $}, we do use the assumption that \Couleur{$ {\bf \delta k}$} is bounded, and we do use (\ref{int |delta p| dS -> 0}).\\

Thus, if \Couleur{$ {\bf k}$} and \Couleur{$\nabla p$} are macro-homogeneous in the sense of assumptions \hyperlink{Ass(i)}{(i)} and \hyperlink{Ass(ii)}{(ii)} respectively, and if \Couleur{$\mathrm{div}\, {\bf k} = 0$}, we may omit the superscript \Couleur{$^{\ \Omega}$} and write
\be\label{NoCorrelation_k_grad_p}
\fbox{$\Couleur{\overline{{\bf k.}\,\nabla p} = \overline{{\bf k}}\,{\bf .}\,\overline{\nabla p}}.$}
\ee

%%%%%%%%%%%%%%%%%%%%%%%%%%%%%%%%%%%%%%%%%%%%%%%%%%%%%%%%%%%%%%%%%%%%%%%%%%%%%%%%
\subsection{Averaged PDE for locally macro-homogeneous fields \Couleur{$ {\bf k}$} and \Couleur{$\nabla p$}}\label{AveragedPDE}
%%%%%%%%%%%%%%%%%%%%%%%%%%%%%%%%%%%%%%%%%%%%%%%%%%%%%%%%%%%%%%%%%%%%%%%%%%%%%%%%

Now we come to the relevant case of locally macro-homogeneous fields, i.e., slow variations of macro-homogeneous fields. In that case, we may use Eq. (\ref{NoCorrelation_k_grad_p}), although the macroscopic volume averages now depend (slowly) on the macroscopic spacetime position. Therefore, the PDE (\ref{1st_order_Eq_for_p}) averages to
\be\label{1st_order_Eq_for_P}
\fbox{$\Couleur{K^\nu P_{,\nu} = F},$} \qquad \mathrm{or}\quad \Couleur{{\bf K.}\nabla P = F}, 
\ee
where 
\be\label{ST-averaged-fields}
\Couleur{{\bf K}:=\overline{{\bf k}},\quad P:=\overline{p}, \quad F:=\overline{f}}.
\ee
That is, Eq. (\ref{1st_order_Eq_for_P}) is the same as (\ref{1st_order_Eq_for_p}), but with spacetime-averaged fields. Those averages have to be taken at a scale where \Couleur{$ {\bf k}$} and \Couleur{$\nabla p$} are (approximately) macro-homogeneous. (Now \Couleur{$R(\Omega)$} cannot be arbitrarily large.) In view of the huge ratio between the galactic scale and the micro-scale (typical wavelength and pseudoperiod), there is enough room.

%%%%%%%%%%%%%%%%%%%%%%%%%%%%%%%%%%%%%%%%%%%%%%%%%%%%%%%%%%%%%%%%%%%%%%%%%%%%%%%%
\section{Solving the PDE for \Couleur{$P$}}\label{Solving-the-PDE}
%%%%%%%%%%%%%%%%%%%%%%%%%%%%%%%%%%%%%%%%%%%%%%%%%%%%%%%%%%%%%%%%%%%%%%%%%%%%%%%%
%%%%%%%%%%%%%%%%%%%%%%%%%%%%%%%%%%%%%%%%%%%%%%%%%%%%%%%%%%%%%%%%%%%%%%%%%%%%%%%%
\subsection{Expression of the solution}
%%%%%%%%%%%%%%%%%%%%%%%%%%%%%%%%%%%%%%%%%%%%%%%%%%%%%%%%%%%%%%%%%%%%%%%%%%%%%%%%

The PDE (\ref{1st_order_Eq_for_P}) for \Couleur{$P$} rewrites as an advection equation
\be\label{advec_P}
\Couleur{\partial _T\, P + U^j \,\partial _j P = S},
\ee
where
\be
\Couleur{S := cF/K^0},\qquad \Couleur{U^j := cK^j/K^0}.
\ee
Therefore, on the characteristic curves
\be\label{cara}
%\fbox{$\Couleur{\frac{\dd {\bf x}}{\dd T} ={\bf U}(T,{\bf x}), \qquad {\bf x}(T_0) = {\bf x}_0},$}
\Couleur{\frac{\dd {\bf x}}{\dd T} ={\bf U}(T,{\bf x}), \qquad {\bf x}(T_0) = {\bf x}_0},
\ee
we have
\be\label{dP/dt}
\Couleur{\frac{\dd P}{\dd T} = \frac{\partial P}{\partial  T} + \frac{\partial P}{\partial  x^j} \frac{\dd x^j}{\dd T} = S(T,{\bf x})},
\ee
so
%\be\label{P on C}
%\fbox{$\Couleur{P(T,{\bf x}(T)) = P(T_0,{\bf x}_0) + \int _{T_0} ^T  S(t,{\bf x}(t))\, \dd t}.$}
%\ee
\be\label{P on C}
\Couleur{P(T,{\bf x}(T)) = P(T_0,{\bf x}_0) + \int _{T_0} ^T  S(t,{\bf x}(t))\, \dd t}.
\ee
Thus, in the particular case that the source field \Couleur{$S$} vanishes, the field \Couleur{$P$} simply is conserved along the characteristic curves (\ref{cara}), which are the trajectories of the continuous medium having velocity field \Couleur{${\bf U}$}.

%%%%%%%%%%%%%%%%%%%%%%%%%%%%%%%%%%%%%%%%%%%%%%%%%%%%%%%%%%%%%%%%%%%%%%%%%%%%%%%%
\subsection{Calculating the micro-field \Couleur{$k^\nu$}}
%%%%%%%%%%%%%%%%%%%%%%%%%%%%%%%%%%%%%%%%%%%%%%%%%%%%%%%%%%%%%%%%%%%%%%%%%%%%%%%%

The 4-vector ``micro-field" \Couleur{$k^\nu$} depends only on the (microscopic) EM field \Couleur{$({\bf E}, {\bf B})$} \cite{A57}:
\bea\label{k^nu}
\Couleur{k^0} & = & \Couleur{\frac{-c}{{\left({\bf E.B}\right)}^2}\, {\bf B.\nabla ({\bf E.B})}},\\
\Couleur{(k^i)} & = & \Couleur{\frac{1}{{\left({\bf E.B}\right)}^2}\, \left(\frac{\partial \left( {\bf E.B}\right)}{\partial T} {\bf B}-{\bf E}\wedge (\nabla ({\bf E.B})) \right )}.
\eea 
To compute \Couleur{$k^\nu$}, we use the ``Maxwell model of the interstellar radiation field", based on axial symmetry (of the galaxy and the ISRF) as a relevant approximation, see Ref. \cite{A63} and references therein. The assumed axisymmetry of the EM field means that its components in cylindrical coordinates \Couleur{$\rho, \, \phi,\, z$} do not depend of the azimuth angle \Couleur{$\phi$}, hence the independent variables are \Couleur{$x^0=t,\,x^1=\rho,\, x^2=z$}. The micro-field \Couleur{$k^\nu $} is thus computed with the said model. That computation uses a set of previously calculated ``spectrum" values \Couleur{$S_{n j}\ (n=0,...,N;\, j=1,..., N_\omega)$}, with here \Couleur{$N=24$} and \Couleur{$N_\omega=23$}. The values \Couleur{$S_{n j}$} depend, in addition to \Couleur{$N$} and \Couleur{$N_\omega$}, on a regular spatial grid used to compute them: here an \Couleur{$(N_\rho = 10,\rho = \rho _0,...,\rho _\mathrm{max})\times (N_z= 21, z=z_0,...,z_{\mathrm{max}} )$} grid with \Couleur{$\rho _0=0, \rho_\mathrm{max}= s_1\times \frac{N_\rho -1}{N_\rho} $}, and \Couleur{$z_0 = -s_1/10, z_{\mathrm{max}} = s_1/10$} \cite{A63}, involving a scale factor \Couleur{$s_1 = 10$} kpc. These numbers mean that the spatial domain used to determine the ``spectrum" values represents the major part of the disk of our galaxy. (That domain is browsed here with quite a rough grid.) Also, \Couleur{$N=24$} for the discretization number of the integration interval for the wave number is a somewhat low value \cite{A63}, chosen here in order to avoid too long times for the computation of the EM field, described just below.\\

Using the \Couleur{$S_{n j}$} values, we calculate the EM field and the associated \Couleur{$k^\nu$} field [Eq. (\ref{k^nu})] on a regular 3D spacetime ``fine" grid of the same form as for the former spatial grid:
\be
\Couleur{x^\mu = x^\mu _0,...,x^\mu_\mathrm{max} \quad \mu = 0,1,2,\ N_\mu} \ \mathrm{values\ for \ \Couleur{x^\mu}}.
\ee
Here this grid had \Couleur{$N_t=40, N_\rho =80, N_z=61,\ t_0= 0,         \ t_\mathrm{max} = 2.4\ \mathrm{kpc}/c,\ \rho_0 = 0.2 \ \mathrm{kpc}, \ \rho_\mathrm{max}=1\ \mathrm{kpc}, \  z_0=0,\ z_\mathrm{max} = 0.061 \ \mathrm{kpc}$},
% t_\mathrm{max} = N_t\delta_t ,\ \rho _0=20 \delta \rho, \rho _\mathrm{max} = \rho _0+N_\rho \delta_\rho,\ z_0=0, \ z_\mathrm{max}= N_z \delta_z $},
% t_\mathrm{max} = (s_2/ c)/0.3 = 2.4 kpc/c
% \rho_0 = 20x0.8/80=20x0.01=0.2 kpc, \rho_\mathrm{max}=rho_0+0.8=1 kpc
% z_\mathrm{max} = 61x(s_2/N_\Rho)/10 = 61x(0.8/80)/10=6.1x0.01=0.061
corresponding with intervals \Couleur{$\delta_t = (s_2/ c)/(0.3 N_t), \delta_\rho = s_2/N_\rho,\ \delta_z = \delta_\rho /10 $} with here \Couleur{$s_2=0.8$} kpc. Thus here a smaller part of the Galaxy is browsed. Moreover, the axis of the Galaxy is avoided, on which the EM field (at least the one predicted by the model) is extremely strong \cite{A63}.

%%%%%%%%%%%%%%%%%%%%%%%%%%%%%%%%%%%%%%%%%%%%%%%%%%%%%%%%%%%%%%%%%%%%%%%%%%%%%%%%
\subsection{The averaged field \Couleur{$K^\nu $} and the superluminal velocity field \Couleur{$U^j$}}
%%%%%%%%%%%%%%%%%%%%%%%%%%%%%%%%%%%%%%%%%%%%%%%%%%%%%%%%%%%%%%%%%%%%%%%%%%%%%%%%

We take the local spacetime average \Couleur{$K^\nu $} of the field \Couleur{$k^\nu $} on a ``rough" grid with steps \Couleur{$\delta_{\mu \,g} = g \delta_\mu$} (with \Couleur{$g$} a small integer). Thus, along each dimension, two successive points of the ``rough" grid are obtained by skipping $g-1$ successive points of the ``fine" grid. The average is done by considering, for each point of the rough grid, its \Couleur{$(2g+1)^3$} nearest neighbours of the fine grid, thus a discrete averaging. E.g. the average takes into account $13^3= 2197 $ points if \Couleur{$g=6$} as considered in the present calculations. \\

\begin{figure}[tbp]
\centering
  \begin{minipage}[b]{0.45\textwidth}
    \includegraphics[width=\textwidth]{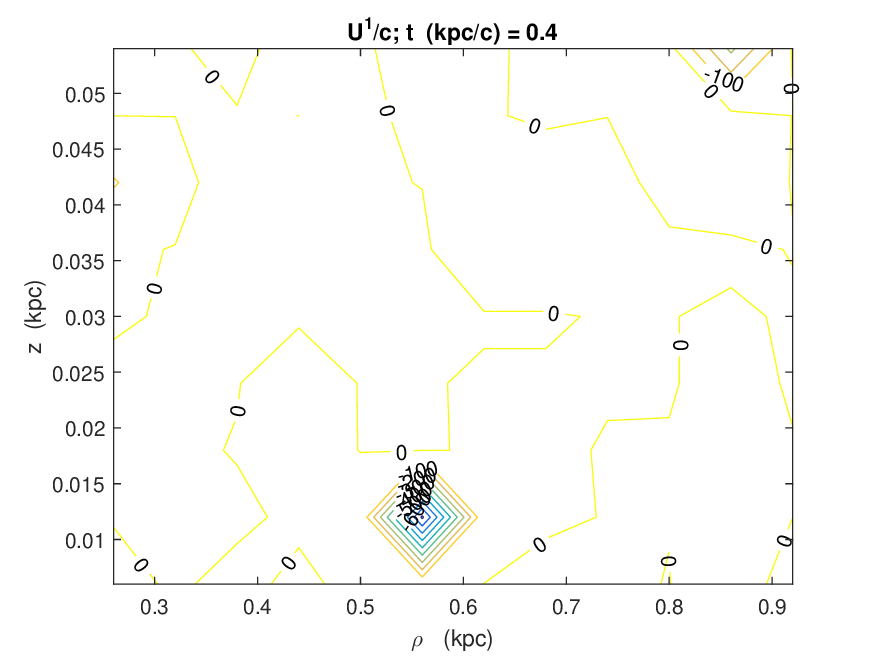}
%    \caption{Flower one.}
  \end{minipage}
  \hfill
  \begin{minipage}[b]{0.45\textwidth}
    \includegraphics[width=\textwidth]{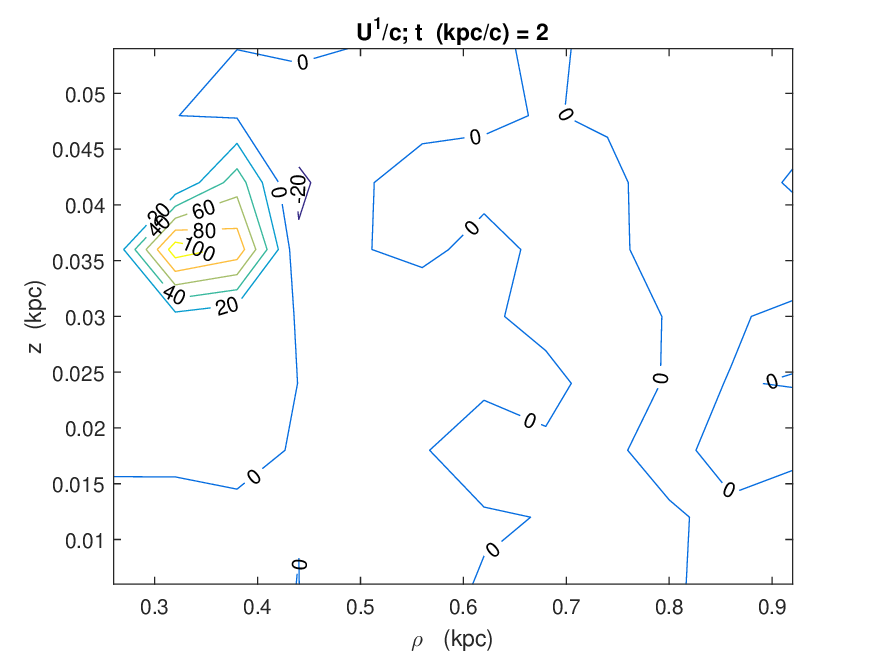}
 %   \caption{Flower two.}
  \end{minipage}
\caption{Component \Couleur{$U^1/c$} at \Couleur{$t = 0.4$} kpc/$c$ or \Couleur{$t=2 $} kpc/$c$.}
\label{U1 g6 t04 & t20}
\end{figure}

\begin{figure}[tbp]
\centering
  \begin{minipage}[b]{0.45\textwidth}
    \includegraphics[width=\textwidth]{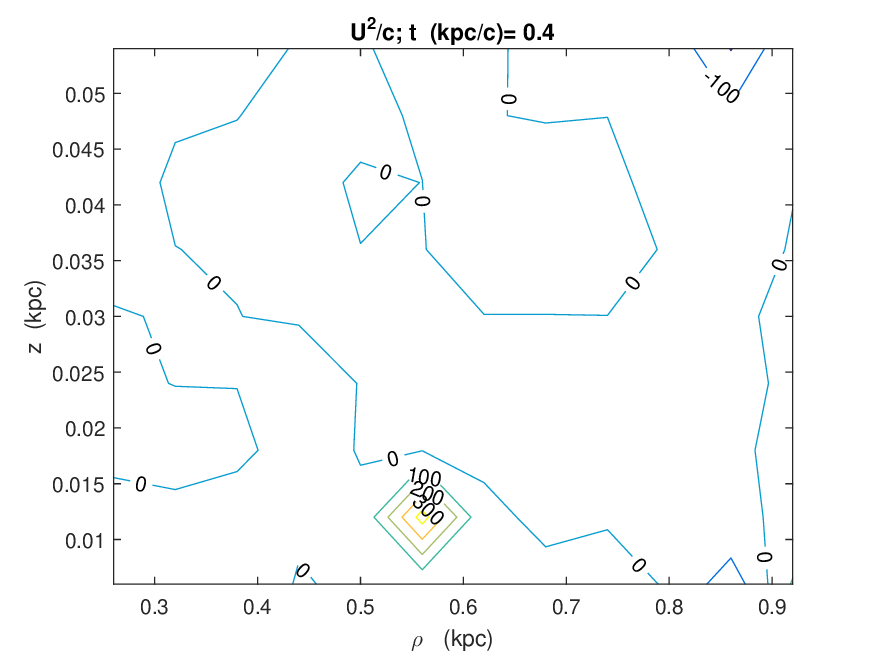}
%    \caption{Flower one.}
  \end{minipage}
  \hfill
  \begin{minipage}[b]{0.45\textwidth}
    \includegraphics[width=\textwidth]{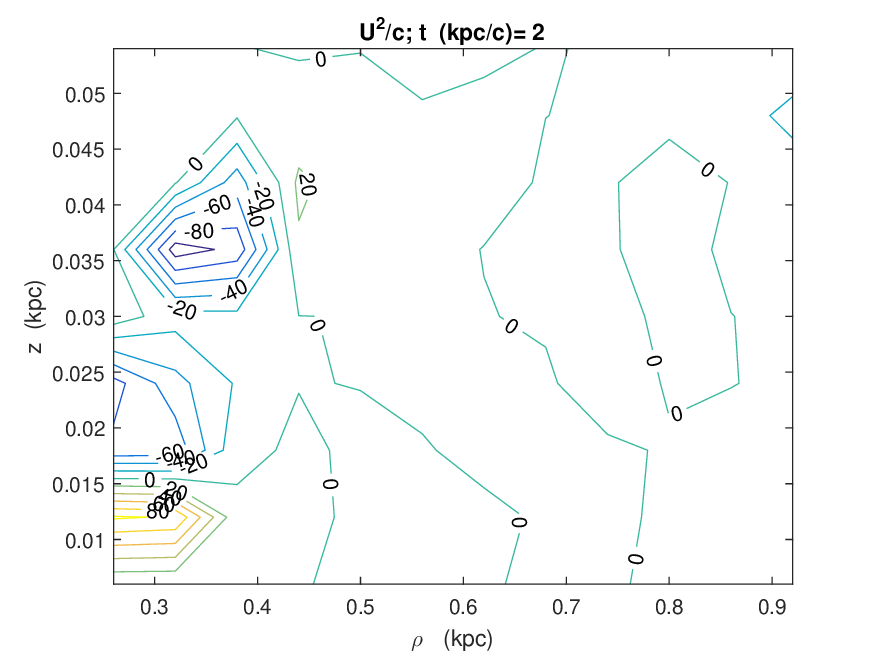}
 %   \caption{Flower two.}
  \end{minipage}
\caption{Component \Couleur{$U^2/c$} at \Couleur{$t = 0.4$} kpc/$c$ or \Couleur{$t=2 $} kpc/$c$.}
\label{U2 g6 t04 & t20}
\end{figure}

\begin{figure}[tbp]
\centering
  \begin{minipage}[b]{0.45\textwidth}
    \includegraphics[width=\textwidth]{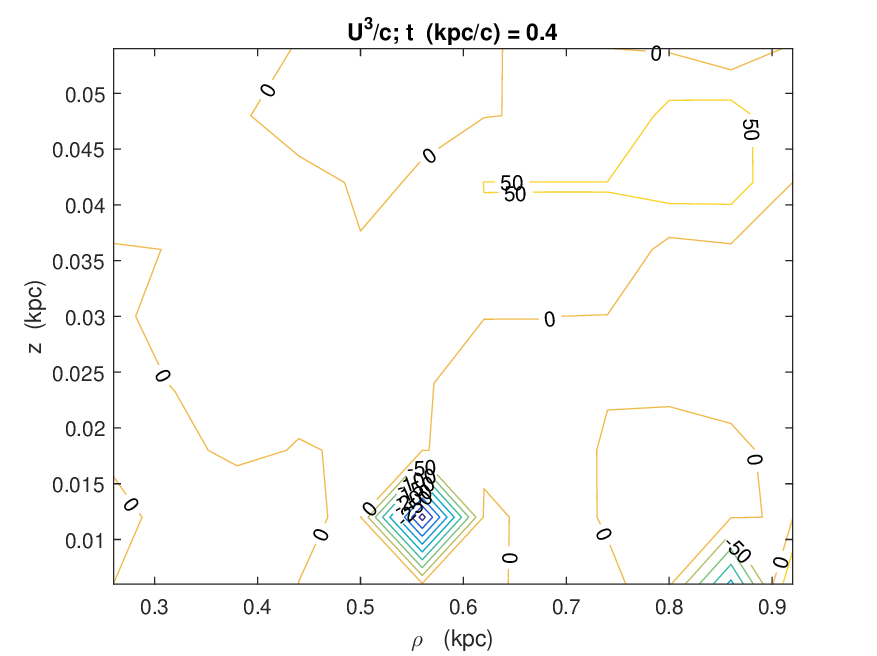}
%    \caption{Flower one.}
  \end{minipage}
  \hfill
  \begin{minipage}[b]{0.45\textwidth}
    \includegraphics[width=\textwidth]{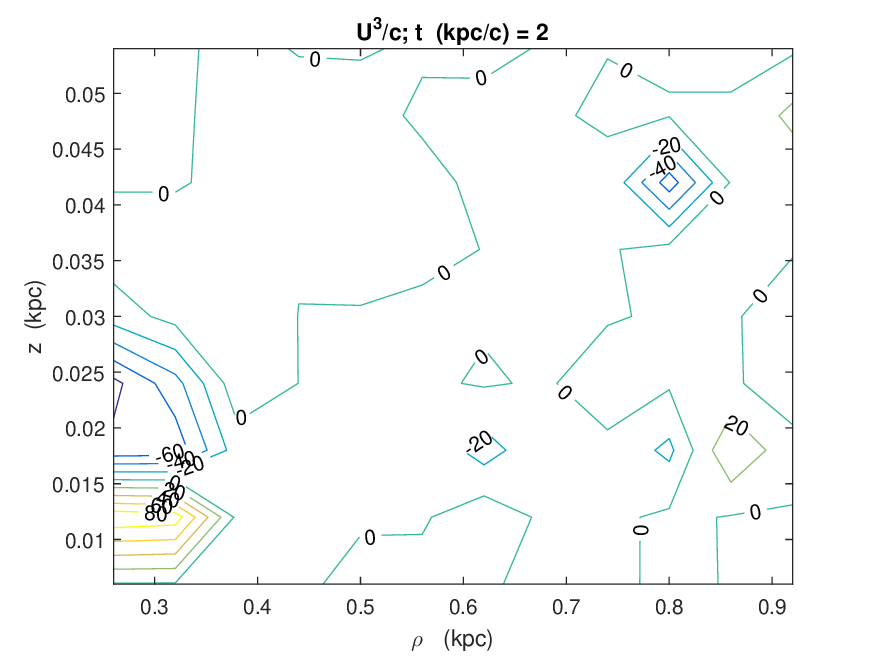}
 %   \caption{Flower two.}
  \end{minipage}
\caption{Component \Couleur{$U^3/c$} at \Couleur{$t = 0.4$} kpc/$c$ or \Couleur{$t=2 $} kpc/$c$.}
\label{U3 g6 t04 & t20}
\end{figure}

Then we calculate the field \Couleur{$U^j = cK^j/K^0$}, whose integral lines are the characteristics (\ref{cara}), and which has clearly the role (and the physical dimension) of a velocity field. Figures \ref{U1 g6 t04 & t20} to \ref{U3 g6 t04 & t20} show contour levels of that velocity field, for the present calculation. We note that it is strongly superluminal in important parts of the domain. According to Eqs. (\ref{cara}) and (\ref{P on C}), the field \Couleur{${\bf U}$} is the velocity field for the ``transport" of the scalar field \Couleur{$P$}. The physical meaning of the field \Couleur{$P$} is that it defines the medium with energy tensor (\ref{T_inter}) (as this medium is seen at the macroscopic scale, due to the average (\ref{ST-averaged-fields})). That medium ``has no rest mass" (it can't be thought of as made of particles with non-zero rest mass), since that energy tensor is very different from the possible energy tensors of any fluid or solid (or EM field). Moreover, as recalled after Eq. (\ref{E_inter}), no velocity can be defined for this medium, and so the field \Couleur{${\bf U}$} is definitely not the velocity field of that medium. Since in SET the upper limit $c$ applies only to the velocity of particles or objects having a non-zero rest mass, we conclude that the superluminal character of the field \Couleur{${\bf U}$} is not a theoretical problem. Also, remind that, in a theory with a preferred reference frame as is SET, there is no causality paradox associated with a superluminal velocity, because the (synchronized) time of the preferred frame is regarded as the ``true time" \cite{O3, A54}.


%%%%%%%%%%%%%%%%%%%%%%%%%%%%%%%%%%%%%%%%%%%%%%%%%%%%%%%%%%%%%%%%%%%%%%%%%%%%%%%
\subsection{Characteristic curves}
%%%%%%%%%%%%%%%%%%%%%%%%%%%%%%%%%%%%%%%%%%%%%%%%%%%%%%%%%%%%%%%%%%%%%%%%%%%%%%%%

Currently we compute some characteristic curves (\ref{cara}). We select two sets of initial conditions: \Couleur{$T= T_0 = 0.8$ kpc/c} for both sets, while the initial positions in the galactic frame \Couleur{$\mathcal{E}_{\bf V}$} (having some relative velocity vector \Couleur{${\bf V}$} with respect to the ether frame \Couleur{$\mathcal{E}$})
\footnote{\ 
Due to the axial symmetry, it is enough to know the modulus \Couleur{$V = \abs{{\bf V}}$} and the angle \Couleur{$\theta $} of \Couleur{${\bf V}$} with the galactic plane.
}
are such that, from one characteristic curve to the other, either:

\bi
\item (i) \Couleur{$\phi_0$} varies and \Couleur{$z'_0=Z_0$} is fixed:
\be\label{fi0 variable}
\Couleur{x'_0= \rho_0\,\cos(\phi_0),\ y'_0= \rho_0\,\sin(\phi_0)}, \quad \Couleur{\phi_0 = k\times \frac{2\pi}{N_\phi}},\quad \Couleur{k =0,...,N_\phi-1},
\ee
or
\item (ii)  \Couleur{$z'_0$} varies and \Couleur{$\phi_0=0$} is fixed:
\be\label{z0 variable}
\Couleur{x'_0= \rho_0, \  y'_0= 0,\ z'_0 = Z_{0 0}+k \delta Z, \quad k=0,...,N_Z-1}.
\ee

\ei

\begin{figure}[tbp]
\centering
  \begin{minipage}[b]{0.45\textwidth}
    \includegraphics[width=\textwidth]{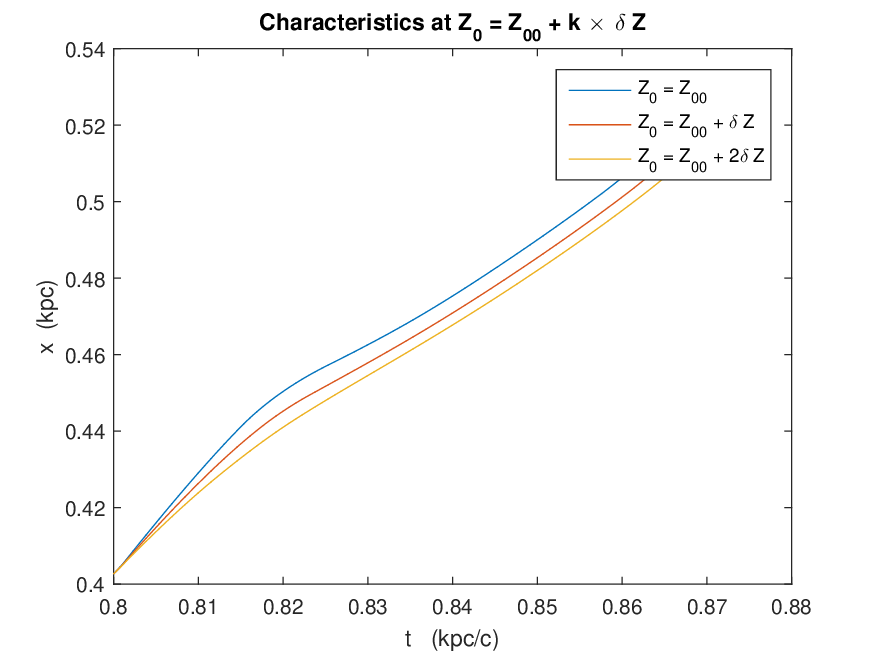}
  \end{minipage}
  \hfill
  \begin{minipage}[b]{0.45\textwidth}
    \includegraphics[width=\textwidth]{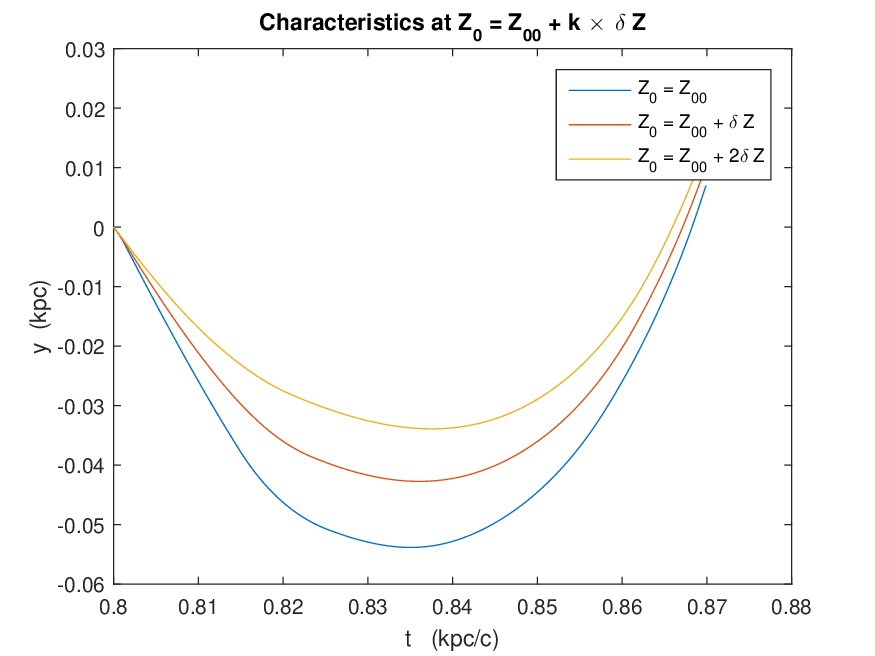}
 %   \caption{Flower two.}
  \end{minipage}
\caption{Three characteristics, set (ii) (see text for details): \Couleur{$(t-x)$} and \Couleur{$(t-y)$} projections.}
\label{Cara g6 tx & ty}
\end{figure}

\begin{figure}[tbp]
\centering
  \begin{minipage}[b]{0.45\textwidth}
    \includegraphics[width=\textwidth]{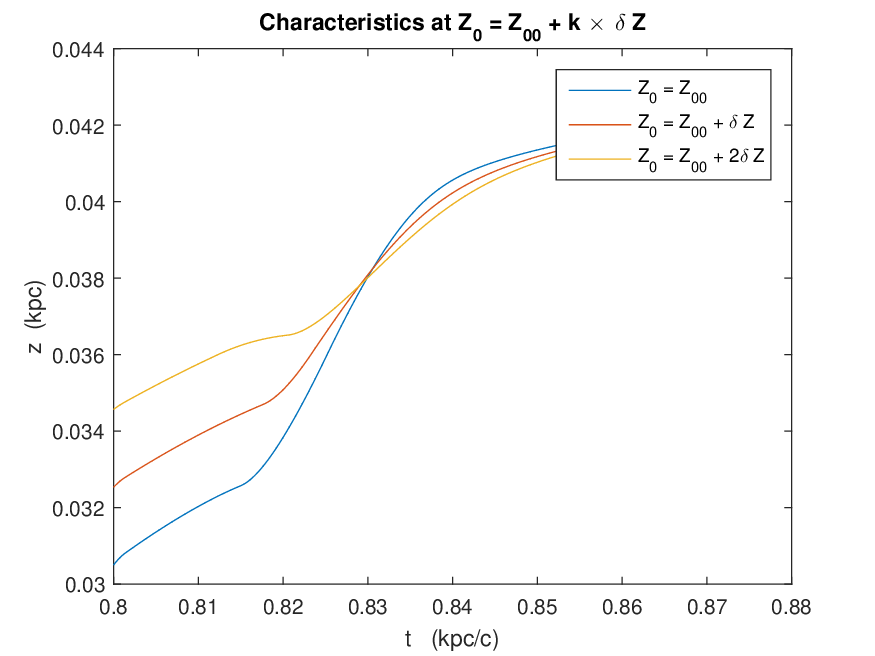}
%    \caption{Flower one.}
  \end{minipage}
  \hfill
  \begin{minipage}[b]{0.45\textwidth}
    \includegraphics[width=\textwidth]{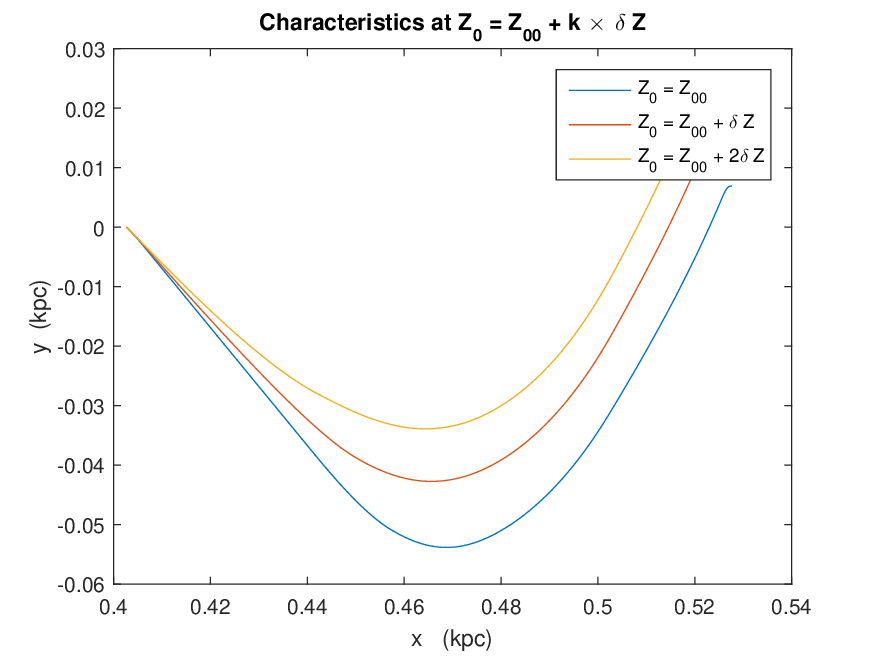}
 %   \caption{Flower two.}
  \end{minipage}
\caption{Three characteristics, set (ii) (see text for details): \Couleur{$(t-z)$} and \Couleur{$(x-y)$} projections.}
\label{Cara g6 tz & xy}
\end{figure}

\begin{figure}[tbp]
\centering
  \begin{minipage}[b]{0.45\textwidth}
    \includegraphics[width=\textwidth]{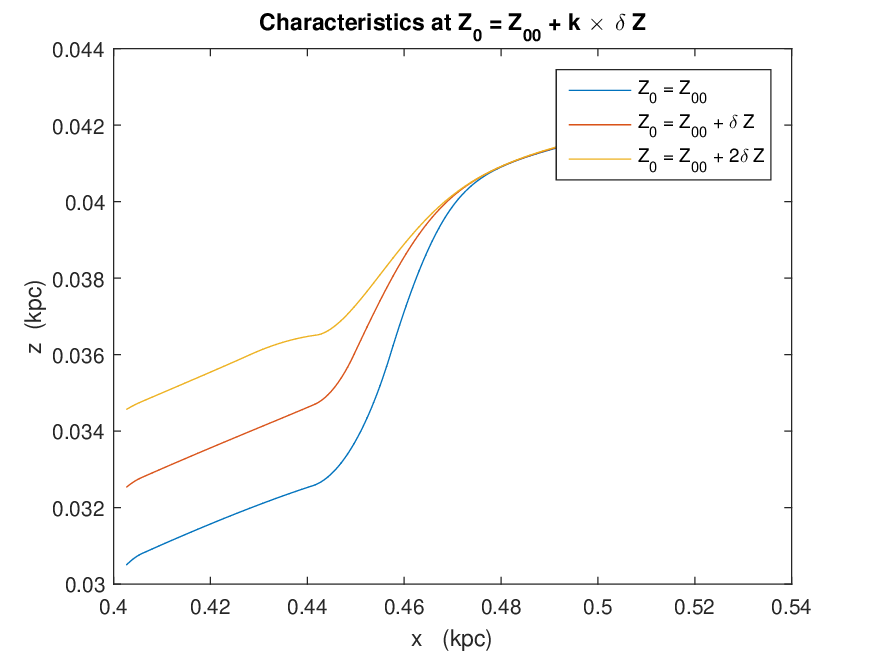}
%    \caption{Flower one.}
  \end{minipage}
  \hfill
  \begin{minipage}[b]{0.45\textwidth}
    \includegraphics[width=\textwidth]{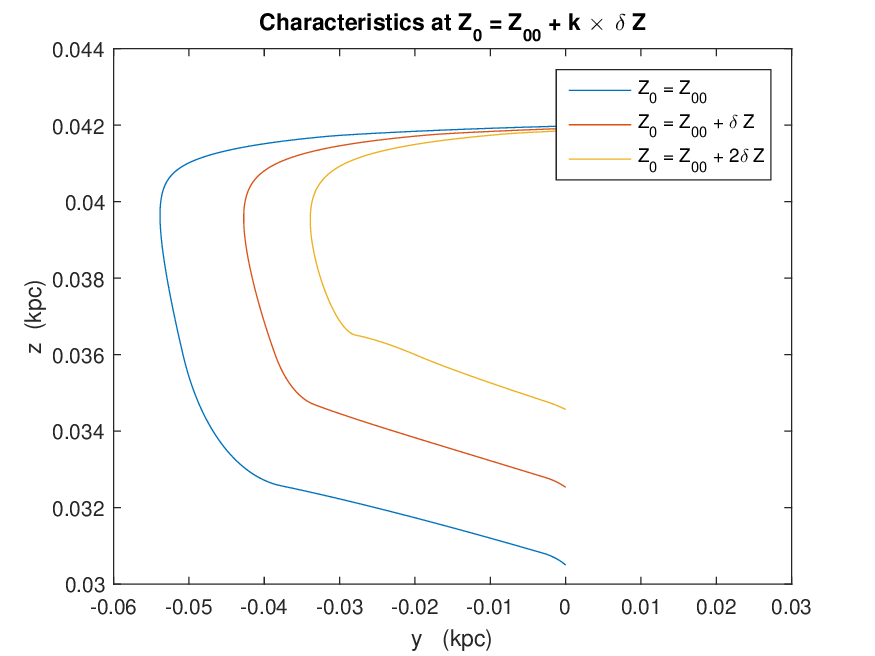}
 %   \caption{Flower two.}
  \end{minipage}
\caption{Three characteristics, set (ii) (see text for details): \Couleur{$(x-z)$} and \Couleur{$(y-z)$} projections.}
\label{Cara g6 xz & yz}
\end{figure}

In both cases, the positions \Couleur{$x_0, y_0, z_0$} in the ether frame \Couleur{$\mathcal{E}$} are obtained by Lorentz transformation, imposing that the time of \Couleur{$\mathcal{E}$} is \Couleur{$T=T_0$}. We then numerically integrate the ODE (\ref{cara}) for the characteristic curves, using the Matlab routine \verb+ode23s+. Figures \ref{Cara g6 tx & ty} to \ref{Cara g6 xz & yz} show the various projections of three characteristic curves, obtained with the second set (\ref{z0 variable}) of initial conditions with \Couleur{$N_Z=3$}. We took \Couleur{$\rho_0 = 0.2000$} kpc, %$\rho _\mathrm{max}= 1$ kpc,\\
\Couleur{$Z_{0 0}= 0.0305$} kpc, \Couleur{$\delta Z= 0.02$} kpc. %$z_0=0$. $z_\mathrm{max}= 0.0610$ kpc with $ Z_{0 0} = (z_\mathrm{max}-z_0)/2$ and $\delta Z=(z_\mathrm{max}-z_0)/(10*NZ)$.

%%%%%%%%%%%%%%%%%%%%%%%%%%%%%%%%%%%%%%%%%%%%%%%%%%%%%%%%%%%%%%%%%%%%%%%%%%%%%%%%
\section{Conclusion}\label{Conclusion}
%%%%%%%%%%%%%%%%%%%%%%%%%%%%%%%%%%%%%%%%%%%%%%%%%%%%%%%%%%%%%%%%%%%%%%%%%%%%%%%%

\bi

\item In the alternative gravity theory ``SET", electromagnetism in the presence of gravitation demands to introduce an additional energy tensor \Couleur{$\Mat{T}_\mathrm{inter}$}, depending on a scalar field \Couleur{$p$} \cite{A57}.\\

\item This exotic energy might contribute to dark matter. The first-order PDE (\ref{1st_order_Eq_for_p}) was derived \cite{A57}. It governs the field \Couleur{$p$} in given EM plus gravity fields.\\

\item A model was developed that provides the EM field in a galaxy, see Ref. \cite{A63} and references therein.\\

\item The quick variation of the EM field prevents integration of (\ref{1st_order_Eq_for_p}) on the scale of a galaxy, so an appropriate averaging has to be found. Several trials with the homogenization theory were not found to provide tractable results; see Ref. \cite{B43} and see subsect. \ref{ST-homogenizn} here.\\

\item Using the theory of macro-homogeneous fields, it was proved here that the same PDE (\ref{1st_order_Eq_for_p}) applies to the macroscopic fields obtained by local spacetime averaging, Eq. (\ref{1st_order_Eq_for_P}). Moreover, we have numerically implemented that local spacetime averaging, by using a discrete average on the points of a finer grid than the one where the macro-fields are sought. \\

\item The PDE (\ref{1st_order_Eq_for_P}) may be rewritten in the form of the advection equation (\ref{advec_P}), that can be integrated along the characteristics (\ref{cara}) by Eq. (\ref{P on C}). The spatial vector field \Couleur{${\bf U}$} that enters the definition of the characteristics is a velocity field. We find that it is partly superluminal, and that, according to the physical role of this field, this is allowed in SET. Currently, we are able to compute characteristic lines at sub-kpc scale.\\

\ei

%%%%%%%%%%%%%%%%%%%%%%%%%%%%%%%%%%%%%%%%%%%%%%%%%%%%%%%%%%%%%%%%%%%%%%%%%%%%%%%%

\section*{References}

\begin{thebibliography}{9}
\small

\bibitem{Einstein1905} Einstein A 1905 Zur Elektrodynamik bewegter K\"orper {\it Ann. der Phys.} (4) {\bf 17} 891-921

\bibitem{Minkowski1907} Minkowski H 1908 Die Grundgleichungen f\"ur die elektromagnetischen Vorg\"ange in bewegten K\"orpern {\it Nachr. Kgl. Ges. d. Wiss. zu Goettingen, Math.-phys. Kl.} {\bf 1908} 53-111

\bibitem{Lorentz1904} Lorentz H A 1904 Electromagnetic phenomena in a system moving with any velocity smaller than that of light {\it Proc. Roy. Netherl. Acad. Arts Sci.} {\bf 6} 809-831

\bibitem{Poincare1905} Poincar\'e H 1905 Sur la dynamique de l'\'electron  {\it C.-R. Acad. Sci. Paris} {\bf 140} 1504-1508

\bibitem{Poincare1906} Poincar\'e H 1906 Sur la dynamique de l'\'electron {\it Rendic. Circ. Matemat. Palermo} {\bf 21} 129-176

\bibitem{Prokhovnik1967} Prokhovnik S J 1967 {\it The Logic of Special Relativity} (Cambridge (U.K.): Cambridge University Press)

\bibitem{Fock1964}
Fock V A 1964 {\it The Theory of Space, Time and Gravitation} second English edition (Oxford: Pergamon)

\bibitem{O3}
Arminjon M 2004 Gravity as Archimedes' thrust and a bifurcation in that theory {\it  Found. Phys.} {\bf 34} 1703-1724

%\bibitem{A16} Arminjon M 1996 On the extension of Newton's second law to theories of gravitation in curved space-time {\it Arch. Mech.} {\bf 48} 551--576

\bibitem{A54} 
Arminjon M 2016 Continuum dynamics and the electromagnetic field in the scalar ether theory of gravitation {\it Open Phys.} {\bf 14} 395-409

\bibitem{A59}
Arminjon M and Winkler R W 2019 Motion of a test particle according to the scalar ether theory of gravitation and application to its celestial mechanics  {\it Z. Naturforsch.} {\bf 74a} 305-316   

%\bibitem{B39} Arminjon M 2017 On continuum dynamics and the electromagnetic field in the scalar ether theory of gravitation {\it J. Phys.: Conf. Ser.} {\bf 845} 012014, 9 pages 

\bibitem{A56} 
Arminjon M 2017 Charge conservation in a gravitational field in the scalar ether theory {\it Open Phys.} {\bf 15} 877-890

\bibitem{A57}
Arminjon M 2018 On the equations of electrodynamics in a flat or curved spacetime and a possible interaction energy {\it Open Phys.} {\bf 16} 488-498

%\bibitem{A58} Arminjon M 2018 Lorentz-invariant second-order tensors and an irreducible set of matrices {\it J. Geom. Sym. Phys.} {\bf 50}, 1-10

%\bibitem{B42} Arminjon M 2023 Interaction energy between a charged medium and its electromagnetic field as a dark matter candidate {\it Proc. Sixteenth Marcel Grossmann Meeting} (Singapore: World Scientific) 2139-2148

\bibitem{B43}
Arminjon M 2023 Towards testing a dark matter candidate that emerges from the scalar ether theory {\it J. Phys.: Conf. Ser.} {\bf 2482} 012021, 11 pages

\bibitem{L&L}
Landau L D and Lifshitz E M 1971 {\it The Classical Theory of Fields} 3rd English edition (Oxford: Pergamon)

\bibitem{Moller1952}
M\o ller C 1952 {\it The Theory of Relativity} (Oxford: Clarendon Press)

\bibitem{Caillerie2012}
Caillerie D 2012 Homog\'en\'eisation des mat\'eriaux \`a structure p\'eriodique, Ecole d'\'et\'e {\it M\'ethodes Asymptotiques en M\'ecanique}, Quiberon, September 2012. Text available at
http://mam.ida.upmc.fr/Telechargements/cours-Caillerie.pdf

\bibitem{A6}
Arminjon M 1991 Macro-homogeneous strain fields with arbitrary local inhomogeneity {\it Arch. Mech.} {\bf 43}, 191-214

%\bibitem{A60}  Arminjon M 2020 An explicit representation for the axisymmetric solutions of the free Maxwell equations {\it Open Phys.} {\bf 18} 255-263

%\bibitem{A61} Arminjon M 2021 An analytical model for the Maxwell radiation field in an axially symmetric galaxy {\it Open Phys.} {\bf 19} 77-90

%\bibitem{A62} Arminjon M 2021 Spectral energy density in an axisymmetric galaxy as predicted by an analytical model for the Maxwell field {\it Adv. Astron.} {\bf 2021}, 5524600, 13 pages

\bibitem{A63} 
Arminjon M 2023 Interstellar radiation as a Maxwell field: improved numerical scheme and application to the spectral energy density {\it Open Phys.} {\bf 21}, 20220253, 17 pages 



\end{thebibliography}
\end{document}